\documentclass[
  aip, jcp,
  reprint,     
  twocolumn, 
  showkeys,    
  showpacs,     
]{revtex4-2}

\usepackage{graphicx}
\usepackage{amsmath,amssymb}
\usepackage[T1]{fontenc}
\usepackage[utf8]{inputenc}   
\usepackage[english]{babel}
\usepackage{nameref}
\usepackage[colorlinks=true,linkcolor=blue,citecolor=blue,urlcolor=blue]{hyperref}

\begin{document}

\title{Deep-Learning Atomistic Semi-empirical Pseudopotential Model for Nanomaterials}

\author{Kailai Lin}
\email{tommy_lin@berkeley.edu}
\affiliation{Department of Chemistry, University of California, Berkeley, California 94720, USA}
\affiliation{Materials Sciences Division, Lawrence Berkeley National Laboratory, Berkeley, California 94720, USA}

\author{Matthew J. Coley-O'Rourke}
\affiliation{Department of Chemistry, University of California, Berkeley, California 94720, USA}

\author{Eran Rabani}
\email{eran.rabani@berkeley.edu}
\affiliation{Department of Chemistry, University of California, Berkeley, California 94720, USA}
\affiliation{Materials Sciences Division, Lawrence Berkeley National Laboratory, Berkeley, California 94720, USA}
\affiliation{The Raymond and Beverly Sackler Center of Computational Molecular and Materials Science, Tel Aviv University, Tel Aviv 69978, Israel}

\begin{abstract}
\noindent\textbf{Abstract}\\
The semi-empirical pseudopotential method (SEPM) has been widely applied to provide computational insights into the electronic structure, photophysics, and charge carrier dynamics of nanoscale materials. We present ``DeepPseudopot'', a machine-learned atomistic pseudopotential model that extends the SEPM framework by combining a flexible neural network representation of the local pseudopotential with parameterized non-local and spin-orbit coupling terms. Trained on bulk quasiparticle band structures and deformation potentials from GW calculations, the model captures many-body and relativistic effects with very high accuracy across diverse semiconducting materials, as illustrated for silicon and group III-V semiconductors. DeepPseudopot's accuracy, efficiency, and transferability make it well-suited for data-driven \textit{in silico} design and discovery of novel optoelectronic nanomaterials.
\end{abstract}

\maketitle


\section{Introduction}
\label{sec:intro}
Semiconductor nanocrystals (NCs) exhibit size-dependent electronic and optical properties that enable their applications in a wide range of technologies.~\cite{alivisatos_perspectives_1996} 
The finite size leads to the discretization of electronic and vibrational states,~\cite{brus_simple_1983} offering tunable fundamental band gaps,~\cite{bawendi_electronic_1990}  optical absorption,~\cite{gomez_optical_2006} Auger lifetimes,~\cite{klimov_quantization_2000} spectral linewidth,~\cite{norris_measurement_1996} and exciton-cooling dynamics~\cite{kumar_hot_2016} that differ drastically from the corresponding bulk materials. Furthermore, engineered nano-heterostructures---such as core-shell NCs,~\cite{reiss_coreshell_2009} alloyed NCs,~\cite{bailey_alloyed_2003} NC arrays,~\cite{hensgens_quantum_2017} and systems with point defects~\cite{babentsov_defects_2008}---can exhibit enhanced or novel properties. The ability to systematically tune composition, size, shape, and heterostructure underscores the potential of computational screening to accelerate the design and discovery of NC materials with tailored properties. Therefore, it is crucial to develop computational methods that can accurately describe the quantum properties of emerging NC materials of experimentally relevant sizes and shapes---including quasiparticle electronic structure, optical excitations, and electron-phonon couplings---within feasible computational cost. 

First-principles methods such as density functional theory (DFT) are widely used to study the electronic structure of materials.~\cite{jain_computational_2016, lejaeghere_reproducibility_2016} However, due to the limitations of approximate exchange-correlation functionals, DFT often underestimates band gaps and yields inaccurate excitation energies compared to experimental measurements.~\cite{cohen_challenges_2012} Several approaches have been used to partially correct these deficiencies at the mean-field level, including hybrid exchange-correlation functionals, the modified Becke-Johnson mBJ functional,~\cite{tran_accurate_2009} and the DFT + U method.~\cite{dudarev_surface_1997, vladimir_i_anisimov_first-principles_1997} To address these limitations more systematically, state-of-the-art approaches employ many-body perturbation theory (MBPT) within the GW approximation, which more accurately accounts for electron-electron interactions via the expansion of the self-energy and provides improved predictions of quasiparticle energies and excited states.~\cite{hybertsen_electron_1986, govoni_large_2015, scherpelz_implementation_2016, golze_gw_2019} Spin-orbit coupling can be incorporated into this approach, and optical absorption spectra can be obtained by solving the Bethe-Salpeter Equation (BSE) on top of GW.~\cite{rohlfing_electron-hole_2000, blase_bethesalpeter_2020} 
Despite their improved accuracy, these first-principles methods incurs a high computational cost, rendering them impractical for routine simulations of large NCs in the moderate to weak confinement regimes.~\cite{makkar_review_2021} In particular, GW/BSE calculations have at least quartic scaling with system size, further compounding this challenge.

On the other hand, semi-empirical methods, such as the pseudopotential methods~\cite{chelikowsky_nonlocal_1976, cohen_application_1984, wang_local-density-derived_1995}, tight-binding (TB) models,~\cite{sutton_tight-binding_1988, kwon_transferable_1994}, and Wannier-function-based TB models~\cite{hamann_maximally_2009, gresch_automated_2018} use parametrized Hamiltonians to simplify the electronic structure problem and significantly lower computational cost. The parameters are usually fitted to experimental data or high-level first-principles calculations (such as GW/BSE), offering alternatives for modeling large NC systems with a good balance of computational cost and accuracy.~\cite{wang_electronic_1994} Particularly, the local density-derived semi-empirical pseudopotential method (SEPM) and its variants have seen many fruitful applications in diverse semiconductor NC systems, providing insights into the electronic structure, photophysics, spectroscopy, and charge carrier dynamics.~\cite{wang_pseudopotential_1996, rabani_electronic_1999, jasrasaria_simulations_2022}  A simple functional form that interpolates well across different form factors was used to describe the pseudopotential in reciprocal space, which was then numerically Fourier transformed to obtain the real-space potential, enabling quick adaptation to nanostructures with broken translational symmetry. 

The resulting real-space NC Hamiltonians can be partially diagonalized with reduced computational cost using iterative methods such as filter diagonalization or Lanczos algorithms to target quasiparticle eigenstates near the band edge.~\cite{wood_new_1985, toledo_very_2002, cullum_lanczos_2002} Later developments extended the SEPM to include non-local terms, spin-orbit coupling (SOC), local strain from deformation, and long-range effects.~\cite{fu_local-density-derived_1997, jasrasaria_simulations_2022, weinberg_size-dependent_2023, coley-orourke_intrinsically_2025} Such extensions introduced more parameters in the pseudopotentials and increased the number of first-principles or experimental properties for the ``fitting'' procedure, such as spinor band structure, effective masses, deformation potentials, and electron-phonon coupling tensors. 


Meanwhile, machine learning models have become popular in computational material science research.~\cite{butler_machine_2018, schmidt_recent_2019, keith_combining_2021, choudhary_recent_2022} By leveraging symmetry preservation and flexible function approximators---such as neural networks, kernel methods and graph neural networks---these models learn an accurate representation of atomic interactions and can achieve near ab initio accuracy with a fraction of the computational cost.~\cite{bartok_machine_2017, schutt_schnet_2018, musil_physics-inspired_2021, deringer_gaussian_2021} Many of these models take the form of machine-learned interatomic potentials (MLIPs),~\cite{behler_generalized_2007, zhang_deep_2018, batzner_e3-equivariant_2022, zeng_deepmd-kit_2023, batatia_mace_2023, kim_learning_2024, cheng_latent_2025} which approximate the Born-Oppenheimer potential energy surface (PES) and enable efficient, transferable predictions of total energies and atomic forces. On the other hand, ML has also been integrated into electronic structure predictions by learning diverse representations~\cite{kulik_roadmap_2022}---including mean-field Hamiltonians,~\cite{schutt_unifying_2019, westermayr_physically_2021, li_deep-learning_2022} electron densities,~\cite{grisafi_transferable_2019, brockherde_bypassing_2017} many-body Green's functions,~\cite{venturella_unified_2025} transferable pseudopotentials,~\cite{woo_neural_2022, kim_transferable_2024, kang_electronic_2025} and tight-binding models~\cite{wang_machine_2021, schattauer_machine_2022}---enabling efficient simulation of the electronic structure in large and complex systems. However, it remains an active area of development to extend these models to describe more complex photophysical phenomena, accounting for non-local correlations, relativistic effects, and deformations resulting from electron-phonon couplings. Previous literature has mainly relied on DFT as the source of training data, which limits the ability to accurately capture band-edge physics and optical properties in semiconducting materials. To address this, incorporating training data beyond the density functional approximation, such as MBPT in the GW approximation, offers a promising path forward. In this post-DFT regime, training directly on quasiparticle eigenenergies becomes essential, as it circumvents the need to approximate inherently non-local, frequency-dependent self-energy terms with effective local potentials or densities.

In this work, we explore the use of machine learning techniques to parametrize semi-empirical pseudopotentials, facilitating its adaptation for computing the electronic, optical, and dynamical properties of novel nanomaterials. We developed a transferable deep-learning atomistic pseudopotential surrogate model (named ``DeepPseudopot''), which combines a neural network local pseudopotential that captures local screened interactions, a non-local angular momentum-dependent correction term, and a spin orbit coupling term to accurately reproduce electronic properties of extended bulk systems. We leverage the flexibility of neural network architectures and the universal approximation theorem~\cite{hornik_multilayer_1989} to model the local pseudopotential in reciprocal space. Combined with the non-local and spin-orbit coupling terms, our model can accurately capture the pseudo-core potential with high precision and incorporate many-body electron interactions beyond the density functional approximation. The DeepPseudopot model is trained to reproduce bulk band structure energies across densely sampled high-symmetry paths in the Brillouin zones and hydrostatic volume deformation potentials obtained from DFT+GW calculations of known lattice phases of semiconductor materials. The model parameters---including the weights and biases of the neural network---are flexibly updated using the backpropagation algorithm, which significantly accelerates the fitting process compared to using a simple functional form and enables locating better fits with lower mean squared error (MSE). We demonstrate that properties like interband transition energies, effective masses, band dispersion, and deformation potentials are accurately captured by our DeepPseudopot model in example systems like Si and group III-V semiconductors. We also show that the resulting atomistic pseudopotentials are transferable, enabling efficient quasiparticle electronic structure calculations for large nanoclusters, alloyed bulk and confined systems, and point defects at the DFT+GW level of theory with significantly lower computational cost. The DeepPseudopot model also integrates seamlessly with existing methods that further compute coupled electron-hole excitations, optical absorption spectra, electron-phonon coupling, and charge carrier dynamics, thereby enabling high-throughput exploration and design of nanomaterials.

This paper is structured as follows. In Section~\ref{sec:II}, we describe the deep-learning pseudopotential (``DeepPseudopot'') Hamiltonian and give details of the model training workflow including data preparation, loss function construction, and model parameter optimization. In Section~\ref{sec:III}, we demonstrate the DeepPseudopot model on the prototypical Si system. We discuss the advantage of the flexible neural network pseudopotential over simple functional forms of traditional SEPM in better fitting the bulk band energies and increased transferability to various lattice phases that are not present in the training data. In Section~\ref{sec:IV}, we illustrate another example of DeepPseudopot on group III-V semiconductor systems (InAs, InP, GaAs, GaP) and the corresponding nanomaterials. We demonstrate that we can use a transferable atomistic DeepPseudopot model to capture the electron interactions across a class of materials. We also show qualitative agreements with experimental measurements in opto-electronic properties and electron-phonon interactions of both binary III-V NCs as well as in alloyed derivatives. In Section~\ref{sec:discussion}, we conclude and give an outlook for future development of machine-learned pseudopotential methods in solid state physics and nanomaterials science. 

\section{Results}
\subsection{Machine learning pseudopotential model and workflow}
\label{sec:II}
\begin{figure*}
\begin{centering}
\includegraphics[scale=0.75]{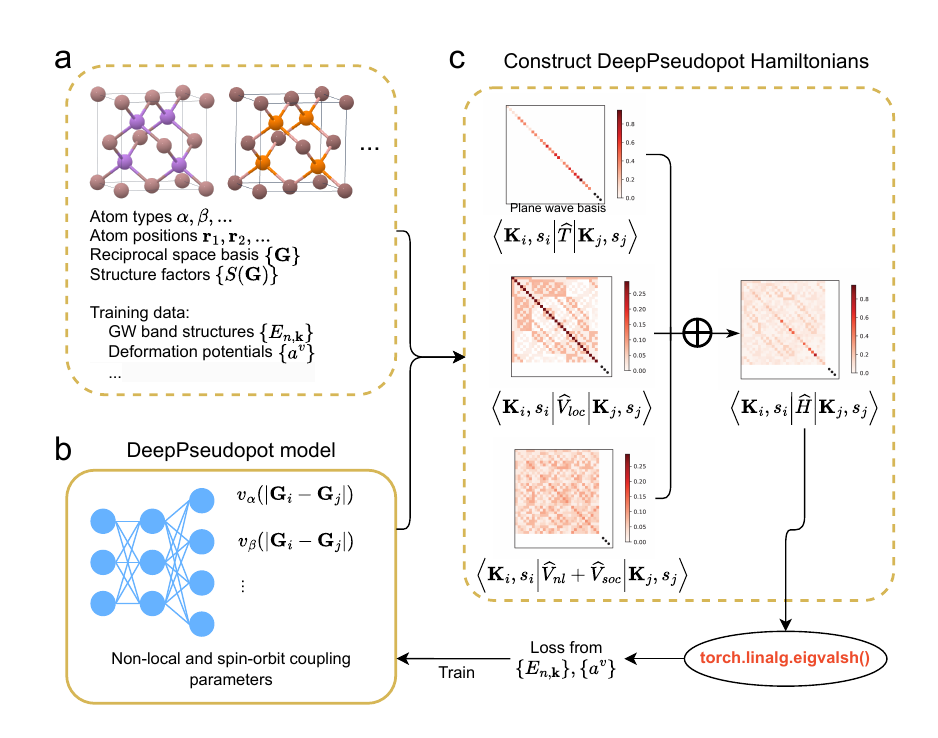}   
\par\end{centering}
\caption{Workflow for developing the DeepPseudopot model. (a) Reference data generation. Quasiparticle band structures and hydrostatic deformation potentials are computed using DFT+GW for multiple crystal structures. (b) Model setup. The atomistic machine learning model is initialized, with the local pseudopotential represented by a neural network and the non-local and spin-orbit coupling terms modeled by parameterized functional forms. (c) Hamiltonian construction and model training. The DeepPseudopot Hamiltonian is constructed from structure factors, wavevector data, and the model pseudopotentials, then diagonalized to obtain the predicted quasiparticle band structures and deformation potentials. The model is trained by minimizing the loss function based on these properties. }
\label{fig:workflow}
\end{figure*}

The machine learning semi-empirical pseudopotential model employs a non-local single-electron Hamiltonian to compute the quasiparticle band structures and deformation potentials in a plane wave spinor basis, as well as the electronic structure of nanoscale systems using a real-space grid basis. The Hamiltonian is given by
\begin{equation}
\hat{H}=\hat{T}+\hat{V}_{\mathrm{loc}}+\hat{V}_{\mathrm{nl}}+\hat{V}_{\mathrm{soc}},
\label{eq::hamil}
\end{equation}
where $\hat{V}_{\mathrm{loc}}$ is the local pseudopotential that acts equally on all angular momentum channels, $\hat{V}_{\mathrm{nl}}$ is an angular momentum-dependent correction to the local pseudopotential, and $\hat{V}_{\mathrm{soc}}$ is the spin-orbit dependent pseudopotential. The pseudopotential terms are given as a sum over atom-centered potentials within the simulation cell, which corresponds to a single unit cell for bulk systems or the full nanocrystal for finite systems. 
\begin{equation}
\hat{V}_{\mathrm{loc}}=\sum_{\alpha}^{\mathrm{N_{at}}}\hat{v}_{\mathrm{loc}}^{\alpha},\quad
\hat{V}_{\mathrm{nl}}=\sum_{\alpha}^{\mathrm{N_{at}}}\hat{v}_{\mathrm{nl}}^{\alpha},\quad
\hat{V}_{\mathrm{soc}}=\sum_{\alpha}^{\mathrm{N_{at}}}\hat{v}_{\mathrm{soc}}^{\alpha}.
\end{equation}
This approach generates the effective potential of a given geometry configuration in a single pass, bypassing the need for computationally intensive self-consistent field (SCF) iterations. 

The local pseudopotential is modeled using a multi-layer fully connected neural network (as illustrated in Fig.~\ref{fig:workflow} and Eq.~\eqref{eq:loc NN}),\cite{hastie01statisticallearning} taking as input the reciprocal space distance $G=\left|\mathbf{G}_{i}-\mathbf{G}_{j}\right|$, where $\mathbf{G}$ denotes the reciprocal space basis. The output tensor size is equal to the number of atom types in the system(s). To enforce the decay of the local pseudopotential in reciprocal space and improve convergence with respect to the kinetic energy cutoff, the activation function in the final layer is replaced with a Gaussian function. The local pseudopotential is given by:
\begin{equation}
v_{\mathrm{loc}}^{\alpha}\left(G\right) = \Bigl[\left(h^{H} \circ \cdots \circ h^{2} \circ h^{1} \right)\left(G\right)\Bigr]_{\alpha}, 
\label{eq:loc NN}
\end{equation}
where $h^{i}(x) = \sigma\bigl(W^{i} x + b^{i}\bigr)$ is the output of the $i$-th hidden layer, $\sigma$ is an activation function, and $W^{i}$ and $b^{i}$ are the weights and bias tensors.\cite{hastie01statisticallearning} In the plane wave spinor basis $\left|\mathbf{K},s\right\rangle $, the local pseudopotential Hamiltonian matrix elements at wavevector $\mathbf{k}$ are expressed as: 
\begin{equation}
\begin{split}
\left\langle \mathbf{K}_{i},s_{i}\left|\hat{v}_{\mathrm{loc}}^{\alpha}\right|\mathbf{K}_{j},s_{j}\right\rangle = \\ \frac{1}{\Omega}e^{i\left(\mathbf{G}_{i}-\mathbf{G}_{j}\right)\cdot\mathbf{R_{\alpha}}}&v_{\mathrm{loc}}^{\alpha}\left(\left|\mathbf{G}_{i}-\mathbf{G}_{j}\right|\right)\delta_{s_{i},s_{j}},
\end{split}
\end{equation}
where $\mathbf{K}=\mathbf{k}+\mathbf{G}$ for all $\mathbf{G}$ of the reciprocal space basis, $e^{i\left(\mathbf{G}_{i}-\mathbf{G}_{j}\right)\cdot\mathbf{R_{\alpha}}}$ is the structure factor $S^{\alpha}\left(G\right)$ for atom $\alpha$, and $\Omega$ is the unit cell volume. We assume spherical symmetry of the local pseudopotential around each atom, which simplifies and accelerates the inverse Fourier transform to real space. The atomic potential does not explicitly encode neighbor information, and is shown to be sufficient for the systems studied here. Environment-dependent descriptors can be systematically incorporated into the model to address more complex systems, as discussed in Section~III. The resulting continuous local potential can be used to construct the nanocrystal potential in the grid basis. Asymmetry in the total electronic potential is captured by the non-local and spin-orbit coupling terms. 

The non-local and SOC pseudopotentials correct the local term, capturing angular-momentum-dependent and relativistic effects. In our model, both terms are assumed to act only on the $l=1$ angular momentum channel (we assume $l=0$ is the local channel) and are represented using simple analytic forms inspired by earlier work.~\cite{chelikowsky_nonlocal_1976, wang_local-density-derived_1995, weinberg_size-dependent_2023, weinberg_expanding_2023}
\begin{align}
\hat{v}_{\mathrm{nl}}^{\alpha}  &=\left[\theta_{\mathrm{nl1}}^{\alpha}e^{-r^{2}}+\theta_{\mathrm{nl2}}^{\alpha}e^{-\left(r-\rho\right)^{2}}\right]\hat{P}_{l=1}^{\alpha},\\
\hat{v}_{\mathrm{soc}}^{\alpha}  &=\left[\theta_{\mathrm{soc}}^{\alpha}e^{-\frac{r^{2}}{w^{2}}}\right]\hat{\mathbf{L}}^{\alpha}\cdot\hat{\mathbf{S}}\,\hat{P}_{l=1}^{\alpha},
\end{align}
where $\hat{P}_{l=1}^{\alpha}$ is the projector onto the $l=1$ orbitals of atom type $\alpha$, $\hat{\mathbf{L}}^{\alpha}$ is the orbital angular momentum operator, $\hat{\mathbf{S}}$ is the spin operator, and $\theta_{\mathrm{nl1}}^{\alpha},\theta_{\mathrm{nl2}}^{\alpha},\theta_{\mathrm{soc}}^{\alpha}$ are the non-local and SOC parameters. We choose to use simple piece-wise Gaussian functions with adjustable prefactors instead of neural networks due to the otherwise prohibitively high computational cost of evaluating and converging the matrix elements in the plane wave spinor basis, given by Eq.~\eqref{eq:SOC matrix}. To further reduce computational overhead, the parameters $\rho$ and $w$ were fixed to $1.5$ and $0.7$ Bohr, respectively. As shown in this work, these approximations---when coupled with the flexibility of the neural network representation of the local pseudopotential---are sufficient to achieve high-quality fits to bulk band structure properties. 

In the plane wave spinor basis, the matrix elements of the non-local and SOC pseudopotentials are expressed as
\begin{widetext}
\begin{equation}
\begin{split}
\left\langle \mathbf{K}_{i},s_{i}\left|\hat{v}_{\mathrm{nl}}^{\alpha}\right|\mathbf{K}_{j},s_{j}\right\rangle  & =\frac{12\pi}{\Omega}\frac{\mathbf{K}_{i}\cdot\mathbf{K}_{j}}{K_{i}K_{j}}e^{i\left(\mathbf{G}_{i}-\mathbf{G}_{j}\right)\cdot\mathbf{R_{\alpha}}}I_{\mathrm{nl}}\,,\\
\left\langle \mathbf{K}_{i},s_{i}\left|\hat{v}_{\mathrm{soc}}^{\alpha}\right|\mathbf{K}_{j},s_{j}\right\rangle  & =-i\frac{12\pi}{\Omega}\frac{\mathbf{K}_{i}\times\mathbf{K}_{j}}{K_{i}K_{j}}e^{i\left(\mathbf{G}_{i}-\mathbf{G}_{j}\right)\cdot\mathbf{R_{\alpha}}}I_{\mathrm{soc}}\cdot\mathbf{S}_{s_{i},s_{j}}\,,
\end{split}
\label{eq:SOC matrix}
\end{equation}
\end{widetext}
where the integrals are defined as $I_{\mathrm{nl}}=\int_{0}^{\infty}dr\:r^{2}\:j_{1}\left(K_{i}r\right)\left(a_{\mathrm{nl1}}^{\alpha}e^{-r^{2}}+a_{\mathrm{nl2}}^{\alpha}e^{-\left(r-\rho\right)^{2}}\right)j_{1}\left(K_{j}r\right)$,
$I_{\mathrm{soc}}=\int_{0}^{\infty}dr\:r^{2}\:j_{1}\left(K_{i}r\right)\left(a_{\mathrm{soc}}^{\alpha}e^{-\frac{r^{2}}{w^{2}}}\right)j_{1}\left(K_{j}r\right)$,
$j_{1}(K_{j}r)$ is the spherical Bessel function of order 1, and $\mathbf{S}_{s_{i},s_{j}}$
are matrix elements of the spin operator. 

Highly accurate reference data from DFT+GW calculations were prepared to train the DeepPseudopot parameters, including the neural network weights and biases for the local pseudopotentials as well as the non-local and SOC parameters. Specifically, we generated band structure data along a densely sampled high-symmetry path in the Brillouin zone and extracted hydrostatic volume deformation potentials. The band structure data trains the pseudopotential model to capture electronic eigenenergies
and their dispersion, while the deformation potentials quantify how the eigenenergies respond to local lattice strains, which is crucial for accurately modeling perturbative electron-phonon coupling. The MBPT correction within GW approximation is essential in this workflow, as it addresses the self-interaction error in standard density functional approximations. Compared to conventional DFT, GW provides significantly more accurate quasiparticle band gaps and excited-state properties, yielding better agreement with experimental measurements in semiconducting materials. Hydrostatic volume deformation potentials describe the change and sensitivity of interband transition energies under isotropic strain: 
\begin{equation}
a_{t}^{V}=\frac{dE_{t}}{d\ln V},
\label{eq:defPot}
\end{equation}
where $E_{t}$ is the transition energy and $V$ is the (deformed) cell volume. To avoid complications associated with the ambiguous absolute energy reference in periodic systems,~\cite{wei_predicted_1999, li_ab_2006} we only calculated deformation potentials for interband transitions. These quantities were evaluated
using a finite difference approach by uniformly expanding and contracting the unit cell and extracting the deformation potential from the slope of the transition energy. This workflow of data preparation from high level of theory can be easily extended to other bulk properties, including electron-phonon coupling tensors, dielectric constants, and charge densities. 

The overall training workflow for the DeepPseudopot model is illustrated
in Fig.~\ref{fig:workflow}. During training, we iterated over the wavevectors $\mathbf{k}$ and constructed the DeepPseudopot Hamiltonian, including the kinetic energy, local pseudopotential, non-local and spin-orbit coupling terms, in a converged plane wave spinor basis. We computed the eigenvalues of the resulting Hamiltonian using the complex Hermitian eigenvalue solver (torch.linalg.eigvalsh()) from PyTorch, which allows for efficient backpropagation through the operation. Degenerate eigenvalues were distinguished by maximizing the overlap of eigenvectors between adjacent $\mathbf{k}$-points, following a Wannier-like process.~\cite{souza_maximally_2001} We also calculated the deformation potential at specified interband transitions via the same finite difference approach. The loss function includes contributions from the band structure (BS) mean-squared error, the deformation potential (defPot) mean-squared error and an optional decay penalty (decay) on the local pseudopotential
\begin{widetext}
\begin{equation}
\mathrm{Loss}=\sum_{n\mathbf{k}}w_{n\mathbf{k}}^{BS}\left(E_{n\mathbf{k}}-\tilde{E}_{n\mathbf{k}}\right)^{2}+\sum_{i}w_{i}^{\mathrm{defPot}}\left(a_{i}^{V}-\tilde{a}_{i}^{V}\right)^{2}+\sum_{\alpha}w_{\alpha}^{\mathrm{decay}}\int dG\,S\left(G\right)\left|v_{loc}^{\alpha}\left(G\right)\right|,
\label{eq:loss}
\end{equation}
\end{widetext}
where $E_{n\mathbf{k}}$ and $\tilde{E}_{n\mathbf{k}}$ are the predicted
and reference eigenvalues for band $n$ at wavevector $\mathbf{k}$.
$a_{i}^{V}$ and $\tilde{a}_{i}^{V}$ are the predicted and reference
deformation potential for transition $i$. $S\left(G\right)=\frac{1}{1+e^{-k\left(G-G_{\mathrm{cut}}\right)}}$
is a shifted sigmoid function that smoothly penalizes non-decaying
components of the local pseudopotential beyond a cutoff momentum.
$w_{n\mathbf{k}}^{BS},w_{i}^{\mathrm{defPot}},w_{\alpha}^{\mathrm{decay}}$
are tunable hyperparameters for balancing the loss function. To emphasize
accurate reproduction of band-edge physics, we used heavy weights
on bands near the conduction band (CB) and valence band (VB) edges,
and on $\mathbf{k}$-points critical to the material's electronic structure. 

\begin{figure*}
\begin{centering}
\includegraphics[width=\textwidth]{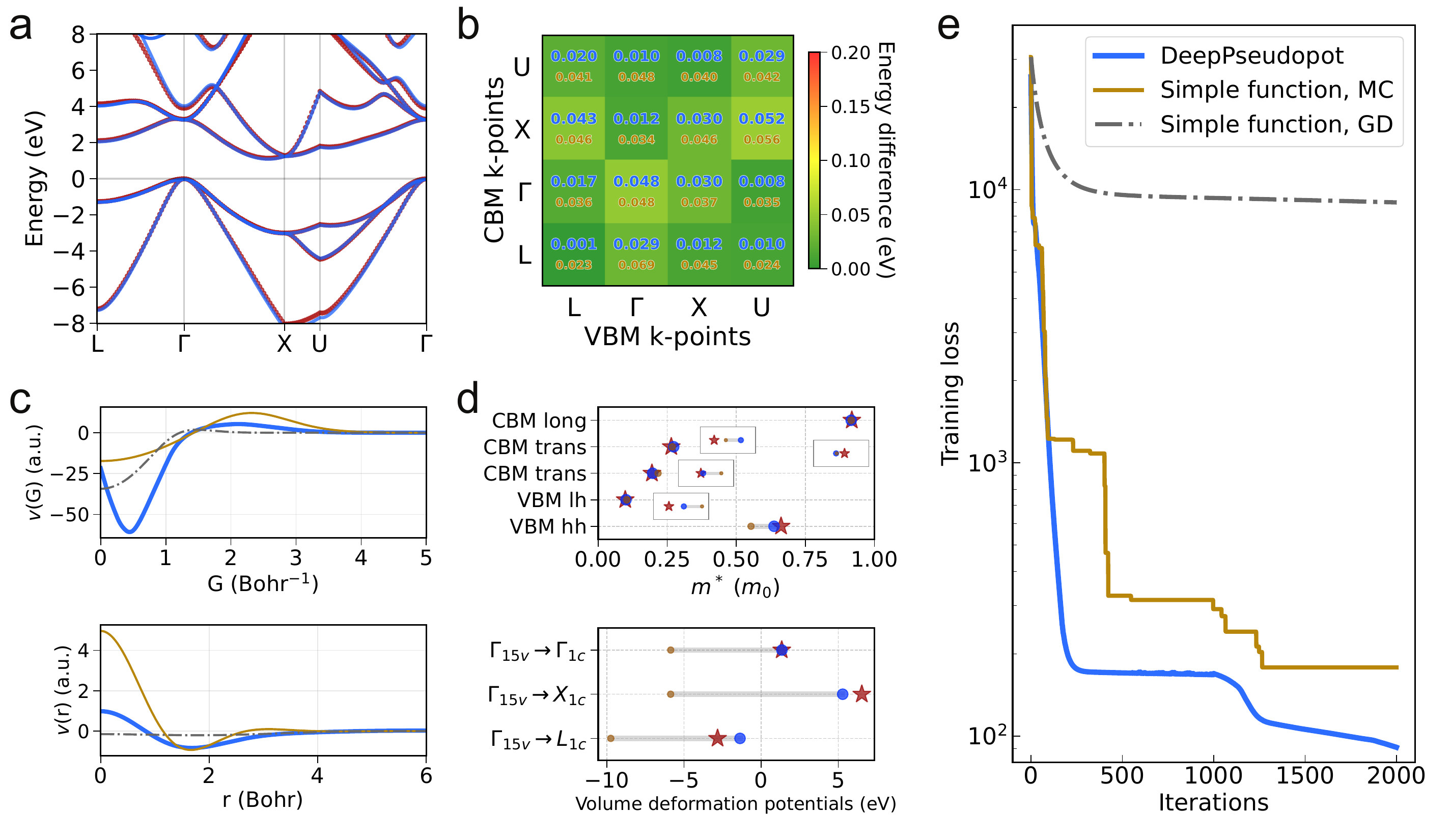}
\par\end{centering}
\caption{Band properties of silicon from training the DeepPseudopot model.
The reference DFT+GW data (red), the DeepPseudopot model predictions
(blue), and the simple functional form pseudopotential fitted using Monte Carlo sampling (yellow) are consistently color-coded across all panels. (a) Band structure
of cubic diamond silicon. (b) Accuracy matrix for
interband transition energies between high-symmetry points in the
Brillouin zone. Grid colors and blue text show the absolute energy errors between reference values and the DeepPseudopot prediction. Yellow text shows the corresponding errors from the simple functional form pseudopotential for comparison. (c) Local pseudopotentials
plotted in reciprocal space and real space. The simple functional form pseudopotential fitted using gradient descent is shown as grey dashed lines. (d) Effective masses (top)
and deformation potentials (bottom). Insets show zoomed-in comparisons of effective masses. (e) Training loss evolution starting from a random initialization. }
\label{fig:siliconPP}
\end{figure*}

\subsection{Application to Si allotropes}
\label{sec:III}
We first demonstrate the versatility and applicability of the DeepPseudopot model using silicon as a test case. The model was trained on DFT+GW reference data for cubic diamond phase silicon, with training and data generation procedures detailed in the Methods section (see Section~\ref{sec:methods}).

The trained silicon machine-learned pseudopotential achieves high accuracy in reproducing reference electronic structure data, as illustrated in Fig.~\ref{fig:siliconPP}. The band structure predicted by the trained DeepPseudopot model closely matches the GW reference along the entire high-symmetry path, accurately capturing both energies and dispersions. To enhance accuracy at the band edges, the training loss function applied double weights to $\mathbf{k}$-points at $\Gamma$, $X$ and at the CBM along the $\Gamma-X$ path. As shown in Fig.~\ref{fig:siliconPP}(b), the resulting model reproduces interband transition energies at high-symmetry points with deviations of less than $0.050$~eV. Notably, it predicts the fundamental band gap with exceptional precision: $1.136$~eV from DeepPseudopot versus $1.137$~eV from DFT+GW. In addition, the model accurately reproduces effective masses and deformation potentials (see Fig.~\ref{fig:siliconPP}(d)), outperforming fits based on simple analytical forms of the local pseudopotential. This high level of agreement indicates that the model can faithfully capture the local electronic potentials, many-body interactions and perturbative properties in the prototypical Si system. 

\begin{figure*}
\begin{centering}
\includegraphics[scale=0.35]{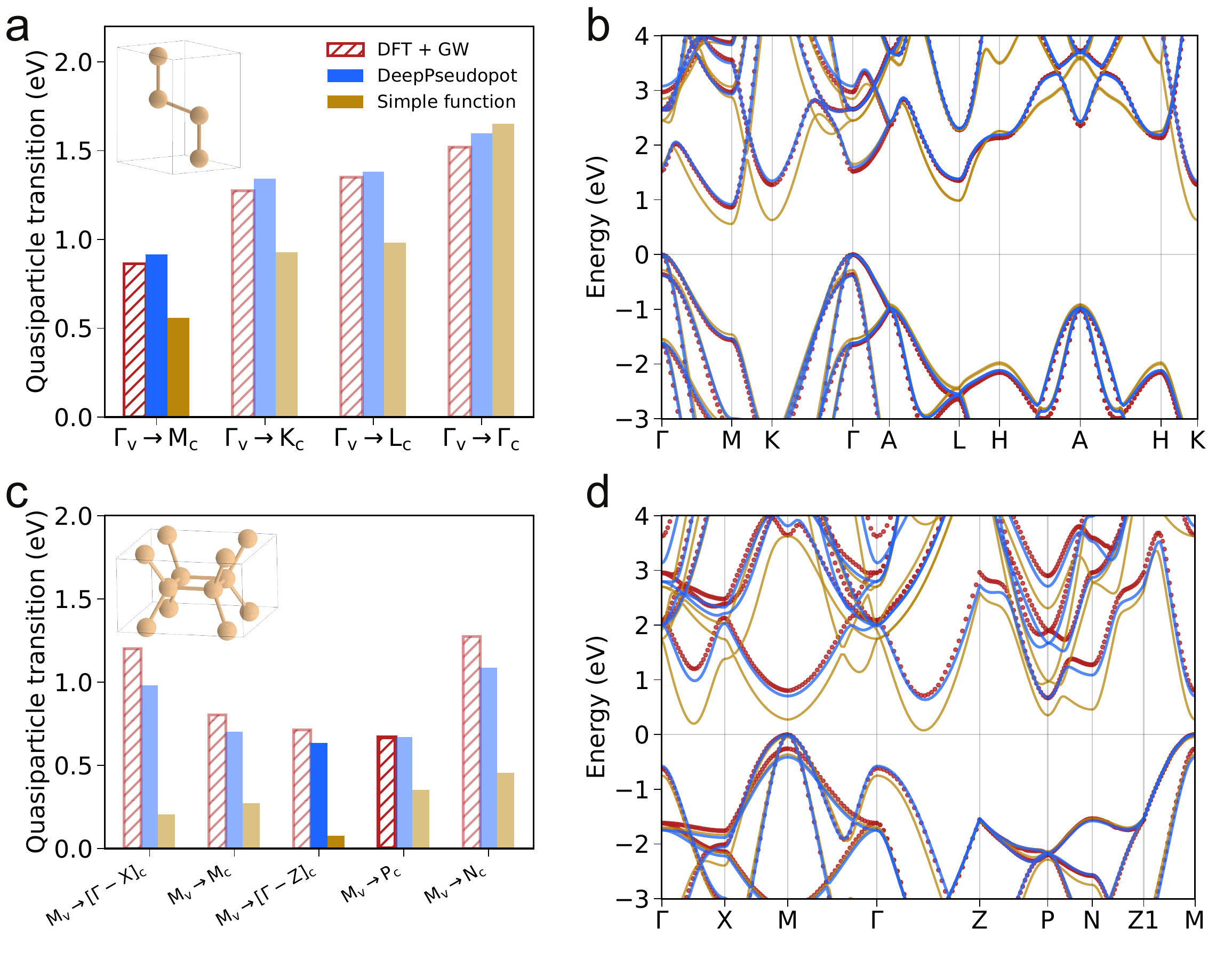}
\par\end{centering}
\caption{DeepPseudopot model predictions for the hexagonal diamond (lonsdaleite) and body-centered tetragonal (bct) structures of silicon. Consistent with Figure 2, reference DFT+GW data are shown in red, DeepPseudopot model predictions in blue, and simple functional form pseudopotentials in yellow. (a) The interband transition energies between the valence band maximum and various $\mathbf{k}$-points of the conduction band edge in the lonsdaleite structure. The fundamental band gaps are highlighted. (b) The reference and predicted band structures of the lonsdaleite structure. (c, d) Same as panels (a) and (b), but for the bct structure.}
\label{fig:silicon}
\end{figure*}

We also compared the training efficiency of the DeepPseudopot model against earlier semi-empirical pseudopotential frameworks.~\cite{wang_local-density-derived_1995, jasrasaria_simulations_2022} Traditional pseudopotentials typically used simple functional forms with only a few tunable parameters, which were adjusted to reproduce band energies. Despite the small parameter space, prior work often relied on stochastic sampling techniques such as Monte Carlo sampling, due to the rugged parameter landscape and the complexity of the eigenvalue operator. To benchmark training performance, we implemented both Monte Carlo (MC) and gradient descent (GD) optimization for the simple functional form, which uses the numerical back-propagation implementation via PyTorch. A common random initialization was selected, with the DeepPseudopot model first trained to reproduce the same initial pseudopotential function as the other two methods to ensure a fair comparison. As depicted in Fig.~\ref{fig:siliconPP}(e), the DeepPseudopot model illustrates improved efficiency, needing fewer iterations of band structure evaluations for a comparable fit. It achieves comparable error levels to MC with less than one-fifth of the computational cost, due to its flexible representation enabling better optimization with respect to the input data.

In contrast, GD on the simple functional form often becomes trapped in a suboptimal local minimum, while MC improves the fit but remains slower and less accurate than DeepPseudopot. Given the empirical nature of the training process and the complexity of the loss landscape, the advantage of DeepPseudopot is robust across runs, but the relative improvement varies with the random initialization. Although demonstrated here for the simple silicon system, the efficiency and flexibility of DeepPseudopot become increasingly important for more complex unit cells. The ability to reach a better fit, reflected in a lower minima of the loss function is critical for achieving the high transferability required of machine-learned pseudopotentials, as demonstrated below.

The trained DeepPseudopot model not only reproduces GW-level reference band properties and deformation potentials at high accuracy within the cubic diamond (cd) phase used for training, but also demonstrates great transferability to other silicon allotropes, as shown in Fig.~\ref{fig:silicon}. To assess the model's predictive performance on unseen structures, we applied it to two additional semiconducting phases of silicon: the hexagonal diamond (lonsdaleite) structure and the body-centered tetragonal (bct) structure, both of which have been studied theoretically or experimentally in literature.~\cite{fujimoto_new_2008, wu_density_2011} While these lattice structures preserve four-fold silicon atom coordination, they exhibit different bond lengths and local atomic environments compared to the cd structure. These structural variations present interesting cases for assessing the model's transferability. 

In both the lonsdaleite and bct structures, the DeepPseudopot model accurately reproduces the GW band dispersions and fundamental band gaps, as shown in Fig.~\ref{fig:silicon}. The predicted transition energies between the VBM and the CB edges at various $\mathbf{k}$-points deviate by less than $0.150$~eV from the GW reference. For comparison, we also computed the band structures using the traditional pseudopotential based on simple analytical functional forms, as trained using MC sampling in Fig.~\ref{fig:siliconPP}(e). Although both models achieved comparable training loss within the cd phase, the more flexible DeepPseudopot model consistently outperforms the simple functional form pseudopotential in prediction tasks on unseen allotropes. The latter severely underestimates the band gaps and misrepresents band dispersions and crossings in the lonsdaleite and bct phases (see Fig.~\ref{fig:silicon}(b) and (d)). 

\begin{figure*}
\begin{centering}
\includegraphics[width=\textwidth]{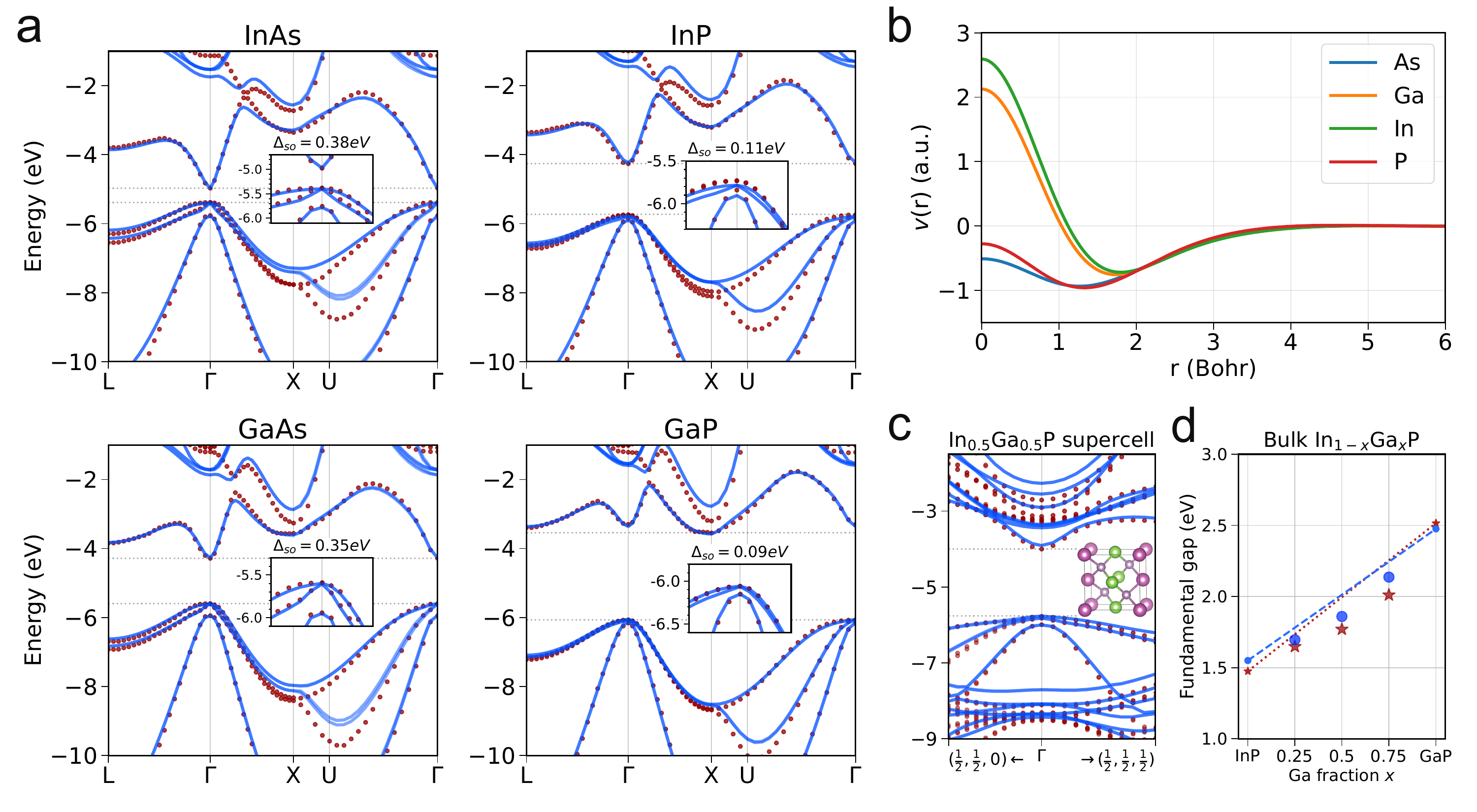}
\par\end{centering}
\caption{Band properties of group III-V semiconductors from training the DeepPseudopot model. Reference DFT+GW data are shown in red throughout; predictions from the trained DeepPseudopot model are shown in blue.
(a) Band structures of InAs, InP, GaAs, GaP, from the trained DeepPseudopot model (blue lines) compared to GW reference (red dots). Insets show zoomed-in views around the VBM, highlighting the spin-orbit splitting energy. (b) Real-space local pseudopotentials for each element. (c) Predicted band structure of alloyed $\mathrm{In}_{0.5}\mathrm{Ga}_{0.5}\mathrm{P}$ around the $\Gamma$ point. The inset shows the 8-atom zincblende conventional cell with cation substitutions used in the calculation. Atom colors: Pink-In, Green-Ga, Purple-P. (d) Fundamental band gaps of bulk $\mathrm{In}_{1-x}\mathrm{Ga}_{x}\mathrm{P}$ alloy supercells.}
\label{fig:III-VPP}
\end{figure*}

However, we note that this level of generalizability without additional retraining or explicit inclusion of atomic environment in the model is not guaranteed across more complex materials and most likely succeeds here due to the relative simplicity of the silicon phases and the similarity between the lonsdaleite, bct, and cd structures. One example where DeepPseudopot shows limited transferability is in the conduction band minimum (CBM) prediction for the bct structure. In the GW reference calculation, the CBM of the bct structure is located at the $P$ point, with a competing minimum along the $\Gamma-Z$ path only $0.044$~eV higher in energy. In contrast, the DeepPseudopot prediction incorrectly identifies the CBM along the $\Gamma-Z$ path (see Fig.~\ref{fig:silicon}(c)), showing the challenges of resolving very small energy differences without explicit retraining. Furthermore, the current model would not be expected to transfer well to amorphous silicon, where the local atomic environments differ significantly, including variations in silicon coordination numbers. To systematically improve transferability, one can expand the training set to include more phases,~\cite{kim_transferable_2024} enabling the flexible machine-learned local pseudopotential to efficiently extrapolate across diverse structural environments.

\subsection{Application to III-V semiconductors and alloyed nanocrystals}
\label{sec:IV}

The SEPM has been widely fitted across multiple crystal structures---such as wurtzite and zinc blende CdSe~\cite{wang_pseudopotential_1996, rabani_electronic_1999}---but its ability to generalize across alloyed systems is less explored. Here, we show how training a DeepPseudopot model on a set of four group III-V semiconductor compounds provides accurate route to the electronic and vibronic properties of binary-compound and ternary-alloyed nanoscale crystal systems in comparison to experimental measurements.

The model was trained on DFT+GW quasiparticle band structures and deformation potentials of InAs, InP, GaAs, and GaP, prepared using the procedures described in the Methods section (see Section~\ref{sec:methods}). Importantly, spin-orbit coupling was explicitly included in the reference calculations, and each band structure was statically shifted to ensure consistent band alignment across materials. To enable transferability to alloyed systems, each elemental pseudopotential (cation or anion) was shared across the two compounds in which the element appears, without interpolation or compound-specific tuning.

As shown in Fig.~\ref{fig:III-VPP}, the trained DeepPseudopot model on group III-V semiconductors accurately reproduces all band-edge properties nearly perfectly across the four materials. To quantify this accuracy, we measured the deviations in quasiparticle transition energies between the VBM and the CB at the $\Gamma$, $X$, $L$ points. This comparison is particularly relevant for predicting alloy behavior since GaP has CBM at $X$, unlike the other three compounds with CBM at $\Gamma$. Moreover, the CB edge at $\Gamma$, $X$, and $L$ in GaP are close in energy, enabling direct-to-indirect gap transitions in III-V alloys involving GaP. The DeepPseudopot model captures these CB energies and the spin-orbit splitting energies within $0.080$~eV, laying a reliable foundation for electronic structure predictions in nanosystems. 
We note that band identity is not explicitly tracked in the training data or model output, as bands are ordered solely by energy at each $\mathbf{k}$-points. As a result, true band crossings may appear visually as avoided crossings in the plotted results, particularly for certain high-energy conduction bands in Fig.~\ref{fig:III-VPP}(a). Despite this, the predicted eigenvectors and corresponding charge densities qualitatively match those from mean-field calculations, further showing that the trained DeepPseudopot accurately captures the underlying physics. In Fig.~\ref{fig:III-VPP}(b), we visualize the learned local pseudopotentials in real space, where clear and physically meaningful similarities emerge between the cation species and between the anion species.

To assess the transferability of the trained DeepPseudopot model beyond its training set, we tested its predictions on bulk III-V alloys. Specifically, we constructed 8-atom zincblende conventional cells of $\mathrm{In}_{1-x}\mathrm{Ga}_x\mathrm{P}$ alloys at simple fractional compositions ($x=0.25, 0.5, 0.75$), which represent the smallest special quasirandom structures commensurate with the stoichiometry.~\cite{zunger_special_1990, wei_electronic_1990} Each alloy supercell was relaxed prior to band structure calculations near the $\Gamma$ point. Despite the absence of explicit In-Ga-P interactions in the training data, the predicted band structures from the DeepPseudopot model closely match the GW reference (see Fig.\ref{fig:III-VPP}c), and the fundamental band gaps across $\mathrm{In}_{1-x}\mathrm{Ga}_x\mathrm{P}$ compositions qualitatively reproduce ab initio trends with very small errors (see Fig.\ref{fig:III-VPP}d). This test on bulk alloys demonstrates the model's ability to generalize beyond the training set of binary semiconductor primitive cells and accurately capture electronic structures of bulk alloy supercells, thanks to its physically motivated Hamiltonian design and elemental sharing parametrization.

\begin{figure*}
\begin{centering}
\includegraphics[width=\textwidth]{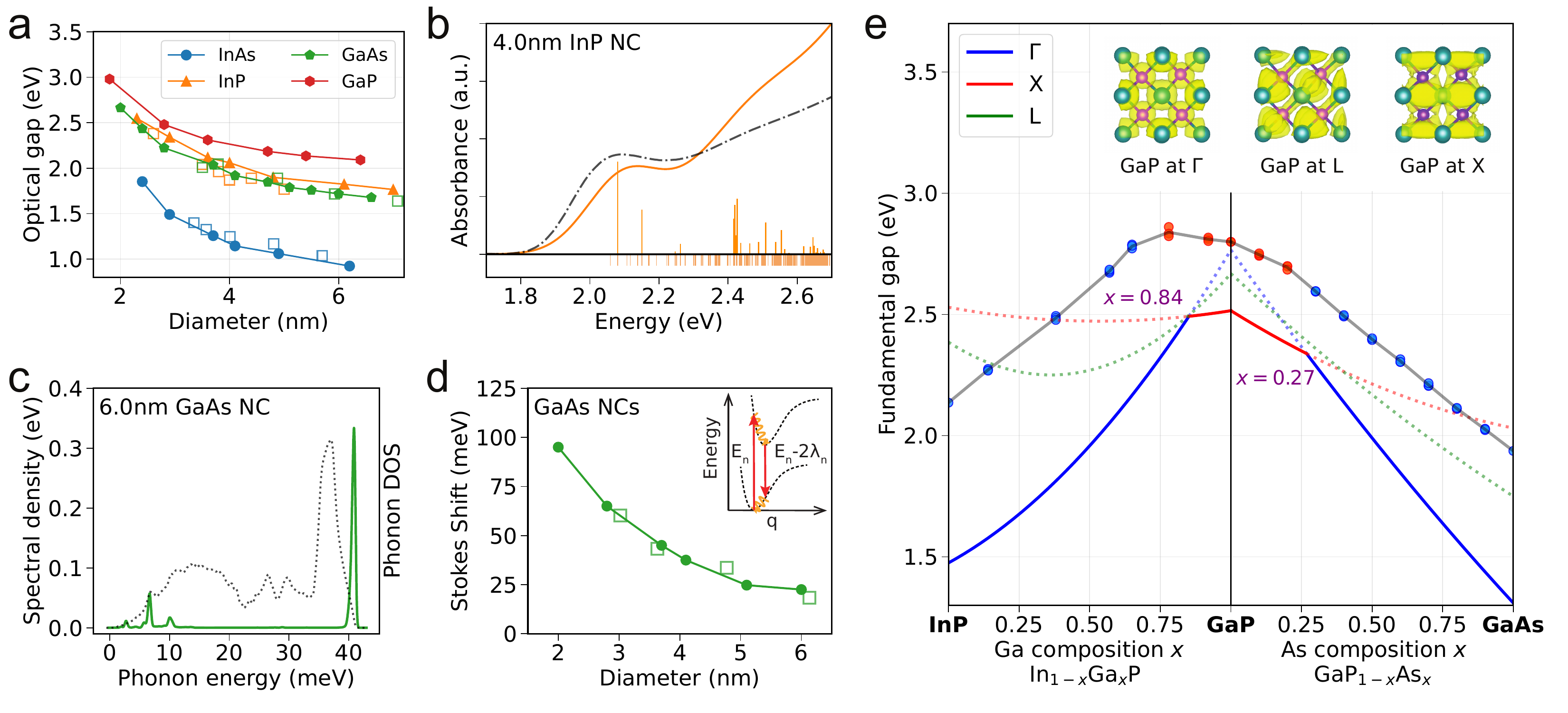}
\par\end{centering}
\caption{Optoelectronic properties and electron-phonon coupling of group III-V NCs calculated using the trained DeepPseudopot model. (a) Size-dependent optical gaps of InAs, InP, GaAs, and GaP NCs compared with experiments (hollow squares).~\cite{micic_synthesis_1994, micic_highly_1996, guzelian_colloidal_1996, ondry_reductive_2024} (b) Calculated (orange solid line) and experimental (black dash-dot) absorption spectra for a $4.0$~nm InP NC. Exciton energies (bars below axis) and oscillator strengths (bars above axis) are shown. (c) Exciton-phonon coupling spectral density (green) and phonon density of states (black dashed line) of a $6.0$~nm GaAs NC. (d) Calculated (solid dots) and experimental (hollow squares) Stokes shifts for GaAs NCs as a function of size. (e) Fundamental band gaps of $\mathrm{In}_{1-x}\mathrm{Ga}_{x}\mathrm{P}$ and $\mathrm{Ga}\mathrm{P}_{1-x}\mathrm{As}_{x}$ ternary alloys. Bulk alloy gaps at the $\Gamma$, $X$, $L$ valleys (dotted lines) were interpolated using a simple quadratic form with experimental bowing parameters~\cite{vurgaftman_band_2001}, with the lowest-energy branch at each composition highlighted as the solid line. Bulk direct-to-indirect crossover compositions are annotated in purple. The NC gaps are shown as dots colored by the dominant GaP valley character (inset). The grey line shows average NC gaps across three random alloy configurations per composition.}
\label{fig:III-V}
\end{figure*}

We then evaluated the predictive capabilities of our trained spinor, non-local DeepPseudopot model for group III-V semiconductors on a variety of nanoscale crystalline systems comprised of the binary III-V compounds, focusing on their optoelectronic properties and electron-phonon coupling. 
Details of the calculations of electronic, optical, and electron–phonon properties in nanocrystals using the trained machine-learned pseudopotentials are provided in the Methods section (see Section~\ref{sec:methods}).

As shown in Fig.~\ref{fig:III-V}, the optoelectronic properties and electron-phonon couplings calculated using the DeepPseudopot model show good agreement with experimental measurements on binary semiconductor nanocrystals. Fig.~\ref{fig:III-V}(a) illustrates the quantum confinement effect for InAs, InP, GaAs, and GaP NCs as a function of size. The optical gaps for all four materials correctly follow the trend of their bulk band gaps, with InAs exhibiting the smallest and GaP the largest gaps. 
The calculated optical gaps also quantitatively agree with experimental measurements, with mean absolute errors of $0.057$~eV for InAs, $0.113$~eV for InP, and $0.046$~eV for GaAs nanocrystals of various sizes.~\cite{micic_synthesis_1994, micic_highly_1996, guzelian_colloidal_1996, ondry_reductive_2024} These results confirm that the DeepPseudopot model captures size-dependent quantum confinement trends and achieves high accuracy in predicting optoelectronic properties across binary III-V nanocrystals.
As a representative example, Fig.~\ref{fig:III-V}(b) shows the computed absorption spectrum of a $4.0$~nm InP NC, where exciton state energies and their OS are represented by bars below and above the axis. The model qualitatively reproduces experimental absorption features, accurately capturing both the first absorption peak position and spectral line shape. 

In Fig.~\ref{fig:III-V}(c) and (d), we further validate the model's exciton-phonon coupling calculations using GaAs NCs. The spectral density calculated using the DeepPseudopot model with phonon density of states (Fig.~\ref{fig:III-V}(c)) obtained from a force field, show structured couplings primarily to a few acoustic phonon modes and strong coupling to discrete optical phonon modes, consistent with prior findings for other semiconductor NCs.~\cite{nomura_excitonlo-phonon_1992, besombes_acoustic_2001, lin_theory_2023} To quantitatively benchmark the overall exciton-phonon coupling strength, we computed the Stokes shift - a collective measure of exciton fine structure and reorganization energy, reflecting overall electron-phonon coupling. As illustrated in Fig.~\ref{fig:III-V}(d), the calculated Stokes shifts for GaAs NCs agree closely with experimental measurements across sizes,~\cite{ondry_reductive_2024} underscoring the model's capability to accurately predict exciton-phonon coupling strengths in nanoscale systems. 

Inspired by recent experimental synthesis development of alloyed III-V NCs in molten salt solvents, we also tested the transferability of the DeepPseudopot model to predict electronic structures of ternary alloyed nanoscale systems. Geometries of $\mathrm{In}_{1-x}\mathrm{Ga}_{x}\mathrm{P}$ and $\mathrm{Ga}\mathrm{P}_{1-x}\mathrm{As}_{x}$ were constructed via random ion exchange starting from pristine tetrahedral InP or GaP NCs, a procedure that mirrors the experimental synthesis pathway.~\cite{gupta_composition-defined_2023} For each alloy composition, we generated three independent, randomly alloyed configurations. These particular alloying systems were chosen due to their intriguing direct-to-indirect band gap transitions involving GaP, which is continuously tunable by adjusting the alloy composition. 

Applying the DeepPseudopot model validated on binary III-V NCs, we predicted the fundamental gaps of ternary alloyed NCs and compared them to theoretical bulk trends. Fig.~\ref{fig:III-V}(e) shows the bulk direct-to-indirect-gap crossover compositions, obtained via quadratic interpolations between binary compounds using GW quasiparticle interband transition energies at $\Gamma$, $X$, $L$ points. Bowing parameters from experimental data were used to account for deviations from linear behavior.~\cite{vurgaftman_band_2001} 
Experimental determination of fundamental gaps in the indirect-gap regime is challenging due to their weak optical emission, making theoretical validation especially valuable. As shown in Fig.~\ref{fig:III-V}(e), the predicted NC fundamental gaps are consistently larger than the bulk values due to quantum confinement, and exhibit nonlinear composition dependence with inflection points closely aligning with bulk crossover compositions. 

In addition, we evaluated the ``majority representation'' coefficients of the CBM states by projecting their quasiparticle wavefunctions onto the bulk GaP Bloch wavefunctions at the direct ($\Gamma$) and indirect ($X$, $L$) valleys (see Fig.~\ref{fig:III-V}(e), inset). Each NC CBM state was classified as either ``$\Gamma$-like'' (blue) or ``$X$-like'' (red). The evolution of these CBM state characters closely tracks the observed linearity changes in the fundamental gap, clearly reflecting the direct-to-indirect-gap transition in $\mathrm{In}_{1-x}\mathrm{Ga}_{x}\mathrm{P}$ and $\mathrm{Ga}\mathrm{P}_{1-x}\mathrm{As}_{x}$ alloyed NCs, despite the broken translational symmetry and the ill-defined nature of quasi-momentum in confined nanoscale systems. 

\section{Discussion}
\label{sec:discussion}
We developed DeepPseudopot, a machine-learning atomistic semi-empirical pseudopotential surrogate model capable of reproducing DFT+GW-level electronic structure properties with very high precision across a diverse set of elemental and compound semiconductors. The model combines a flexible neural network architecture of the local screened pseudopotentials with analytically tractable non-local and spin-orbit coupling terms to capture angular-momentum-dependent and relativistic effects. Physically motivated design choices---including species-specific potential sharing without interpolation, reciprocal-space decay regularization, and targeted loss function weighting at key bands and $\mathbf{k}$-points---enable accurate description of band-edge physics, including quasiparticle energies, deformation potentials, and effective masses. Applied to silicon and group III-V semiconductors (InAs, InP, GaAs, GaP), DeepPseudopot achieves quantitative agreement with GW reference data and significantly outperforms traditional analytic semi-empirical pseudopotentials in both accuracy and training efficiency. The model was shown to generalize well to unseen crystal phases, bulk alloys, large nanostructures, as well as alloyed nanostructures, capturing essential features of electronic, optical, and vibronic properties with no additional retraining. This showcases that the DeepPseudopot model can learn very efficiently from only a small dataset of bulk band structures and deformation potentials, then directly applied to perform transferable, highly accurate electronic structure calculations for large nanoscale systems at a fraction of the computational cost of the reference theory. DeepPseudopot is thus positioned as a broadly applicable tool for the data-driven design and discovery of complex nanomaterials with tailored properties.

Despite accurately describing silicon and III-V alloys, several areas in DeepPseudopot require further development. The non-local and SOC terms are currently confined to one angular momentum channel and are represented by simple analytical functions with limited tunable parameters, a design choice necessitated by the high computational cost of evaluating these terms in the spinor plane wave basis. Developing more efficient algorithms for constructing these matrices could enable higher angular momentum projections and allow for flexible neural network representations. Additionally, the local term assumes spherical symmetry and introducing symmetry-preserving descriptors of the local atomic environment, as suggested by Kim and Son,\cite{kim_transferable_2024} could improve the DeepPseudopot performance. Furthermore, the current local pseudopotential lacks explicit long-range treatment,\cite{cheng_latent_2025} which, while sufficient for capturing bulk deformation potentials and exciton-phonon coupling in nanocrystals, may not accurately model Frohlich-type electron-phonon interactions.\cite{coley-orourke_intrinsically_2025} These developments would broaden DeepPseudopot's applicability to a wider range of materials, and will be the subject of future development. 
\section{Methods}

\label{sec:methods}
\subsection{Machine Learning}
Our proposed machine learning semi-empirical pseudopotential may be conveniently implemented via all common ML packages. We built our model via PyTorch. Model parameters were optimized using the Adam algorithm~\cite{kingma_adam_2017} with an initial learning rate $\beta \approx 0.002$ and an exponential decay scheduler. Training was parallelized over $\mathbf{k}$-points for computational efficiency. The fully trained models were achieved after roughly 2000 epochs for silicon and 15,000 epochs for the III–V systems. A step-by-step model training workflow is illustrated in Fig.~\ref{fig:workflow}. Starting from atomic structures, the model generates the reciprocal-space basis and structure factors, constructs the kinetic, local, non-local, and SOC components of the Hamiltonian, and trains its parameters against reference GW band-structure and deformation-potential data.

For silicon, the local pseudopotential was parameterized by a fully connected neural network with a single hidden layer containing 20 neurons and the Continuously Differentiable Exponential Linear Unit (CELU) activation function,~\cite{barron_continuously_2017} except in the output layer as described in Section~\ref{sec:II}. He initialization~\cite{he_delving_2015} was used to set initial weights. To promote smooth behavior in reciprocal space, a small regularization term was included in the loss function to penalize nonzero components of the local pseudopotential beyond the momentum cutoff of $4.5\mbox{~Bohr}^{-1}$. Spin-orbit coupling and non-local potentials were omitted during training, as SOC effects in silicon are negligible and an accurate band structure could be trained without including non-local angular momentum-dependent corrections.

For the III–V materials, the DeepPseudopot model was constructed with one hidden layer of $50$ neurons and the CELU activation function, with four outputs corresponding to the local pseudopotentials of P, Ga, As, and In. Four accompanying SOC parameters were also included and initialized randomly. 
Each elemental pseudopotential (cation or anion) was shared between the two materials in which the element appears (e.g., In shared between InP and InAs), without any interpolation across systems. This design is particularly important for modeling alloyed systems. To prioritize accurate reproduction of band-edge physics, the loss function used to train the DeepPseudopot model was heavily weighted towards bands near the gap, the spin split-off bands at $\Gamma$ point, and $\Gamma$, $X$ and $L$ points in the Brillouin zone. 

\subsection{Ab-initio bulk band property calculation}
Reference DFT+GW data were generated for silicon and the III–V semiconductors InAs, InP, GaAs, and GaP. These data include quasiparticle band structures and hydrostatic volume deformation potentials for key interband transitions, which were used to train the DeepPseudopot model.

For the mean-field calculations of silicon, we used the Perdew-Burke-Ernzerhof exchange-correlation functional~\cite{perdew_generalized_1996} with norm-conserving, scalar relativistic pseudopotentials as implemented in Quantum ESPRESSO.~\cite{giannozzi_quantum_2009} Spin-orbit coupling effects were neglected, as they are known to be weak in Si. A kinetic energy cutoff of 120 Ry was used, and the Brillouin zone was sampled using a Monkhorst-Pack $8\times8\times8$ $\mathbf{k}$-point mesh in the self-consistent DFT calculation. The GW corrections were performed using the BerkeleyGW package within the single-shot G0W0 approximation.~\cite{deslippe_berkeleygw_2012} The energy cutoff for the screened Coulomb interaction, as well as the number of bands used in the screened Coulomb and Coulomb-hole summations, were converged to ensure numerical accuracy. Additionally, we computed the deformation potentials for interband transitions $\Gamma_{15v}-\Gamma_{1c}$, $\Gamma_{15v}-X_{1c}$, $\Gamma_{15v}-L_{1c}$ at the same level of DFT+GW theory using a unit cell with $\pm1\%$ isotropic deformation of the lattice constant, following Eq.~\eqref{eq:defPot}. 

For InAs, InP, GaAs, and GaP, reference GW calculations were carried out using similar procedures, with several important distinctions. Among these III-V semiconductors, heavier elements such as In and As are known to lead to significant SOC effects. To ensure consistency, fully relativistic pseudopotentials were employed for all elements in the DFT calculations. In the case of InAs, due to the small fundamental gap, DFT calculation using the PBE functional yields a semi-metallic system. Thus, we performed a second iteration of the screened exchange summation using updated G0W0 quasiparticle energies to correctly account for state occupations.~\cite{malone_quasiparticle_2013} Given that the model was trained across multiple compounds and will be applied to nano-heterostructures, consistent band alignment was essential. We statically shifted each band structure so that its VBM aligns with its experimental work function.~\cite{freeouf_schottky_1981} Deformation potentials were computed using the same procedure and included in the training dataset. Bulk alloy band properties were computed using the same procedures, applied to 8-atom zincblende conventional cells with appropriate elemental substitutions. Each alloy cell was first relaxed using DFT with the PBE functional and spin-orbit coupling included, followed by GW  quasiparticle band structure calculations using the same approach described above. 

\subsection{Machine-learned pseudopotential calculation of nanocrystal properties}

The nanocrystal structures were constructed by cutting desired geometries from bulk lattices, followed by structural relaxation using a previously parameterized Tersoff-type force field~\cite{powell_optimized_2007} and surface passivation with ligand potentials.~\cite{wang_pseudopotential_1996} The NC Hamiltonians were constructed using the trained real-space pseudopotentials on a finely spaced real-space spinor grid basis with $0.5$~Bohr spacing, ensuring an accurate representation of the non-local and spin-orbit coupling terms via the projector formalism and the convergence of eigenvalues. The quasiparticle eigenstates near the band edges were efficiently computed using the filter diagonalization method. 

Correlated electron-hole excitations (exciton states) were obtained by solving the Bethe-Salpeter equation within the static screening approximation, using the quasiparticle states obtained from the DeepPseudopot Hamiltonian as the electron-hole product basis. Size-dependent dielectric constants required for BSE calculations were estimated from bulk values using the generalized Penn model.~\cite{williamson_pseudopotential_2000} Oscillator strengths (OS) were calculated from the transition dipole moments between the ground and excitonic states. Phonon frequencies were obtained by diagonalizing the Hessian matrix constructed using a classical Tersoff-type force field for computational efficiency.~\cite{powell_optimized_2007} First-order exciton-phonon couplings were computed via numerical differentiation of the real-space pseudopotentials.~\cite{jasrasaria_interplay_2021} To simulate experimentally measured Stokes shifts, we subtracted the emission peak energy---calculated from exciton energies redshifted by twice the reorganization energy---from the first peak of the absorption spectrum.~\cite{ondry_reductive_2024} More details on the methods for III-V NC construction, BSE, oscillator strength, and Stokes shift calculations can be found in previous work.~\cite{jasrasaria_simulations_2022, gupta_composition-defined_2023}

\section{Data Availability}
The machine-learned pseudopotential parameters, trained and reference band structure data, and nanocrystal simulation results are available on Figshare at \url{https://doi.org/10.6084/m9.figshare.29321645}.

\section{Code Availability}
The DeepPseudopot package is publicly available at \url{https://github.com/TommyLinkl/DeePseudopot.git}

\section{Acknowledgments}
We thank Professors Bingqing Cheng and David Limmer for valuable discussions. This work was supported by the National Science Foundation Division of Chemistry, under the Chemical Theory, Models and Computational Methods (CTMC) program, grant number CHE-2449564. Methods used to describe the vibronic properties of NCs were provided by the center on ``Traversing the death valley separating short and long times in non-equilibrium quantum dynamical simulations of real materials'', which is funded by the U.S. Department of Energy, Office of Science, Office of Advanced Scientific Computing Research and Office of Basic Energy Sciences, Scientific Discovery through Advanced Computing (SciDAC) program, under Award No. DE-SC0022088. Measured optical properties of III-V NCs were supported by the National Science Foundation Science and Technology Center for Integration of Modern Optoelectronic Materials on Demand (IMOD) under award DMR-2019444. This research used resources of the National Energy Research Scientific Computing Center (NERSC), a DOE Office of Science User Facility supported by the Office of Science of the U.S. Department of Energy under Contract No. DE-AC02-05CH11231 using NERSC award BES-ERCAP0032503.

\section{Author Contributions}
K.L. and E.R. conceived and designed the project and co-wrote the manuscript. K.L. developed the machine-learned semi-empirical pseudopotential model and associated codebase, performed ab initio GW calculations, trained and evaluated the model, and conducted data analyses. M.J.C. contributed to model implementation and supported manuscript writing. E.R. supervised the research and acquired funding. All authors discussed the results and contributed to manuscript revisions.

\section{Competing Interests}
The authors declare no competing interests.

\bibliographystyle{naturemag}  
\bibliography{references}
\end{document}